\documentclass[prl,twocolumn,showpacs,floatfix,preprintnumbers,amsmath,amssymb,superscriptaddress]{revtex4-1}

\usepackage{graphicx}
\usepackage{color}
\usepackage[latin1]{inputenc}

\begin{document}

\title{Superconducting gap symmetry in BaFe$_{1.9}$Ni$_{0.1}$As$_{2}$ superconductor}

\author{T.E. Kuzmicheva}
\affiliation{P.N. Lebedev Physical Institute, Russian Academy of Sciences, 119991 Moscow, Russia}

\author{S.A. Kuzmichev}
\affiliation{M.V. Lomonosov Moscow State University, Faculty of Physics, 119991 Moscow, Russia}
\affiliation{P.N. Lebedev Physical Institute, Russian Academy of Sciences, 119991 Moscow, Russia}

\author{A.V. Sadakov}
\affiliation{P.N. Lebedev Physical Institute, Russian Academy of Sciences, 119991 Moscow, Russia}
\author{S.Yu. Gavrilkin}
\affiliation{P.N. Lebedev Physical Institute, Russian Academy of Sciences, 119991 Moscow, Russia}
\author{A.Yu. Tsvetkov}
\affiliation{P.N. Lebedev Physical Institute, Russian Academy of Sciences, 119991 Moscow, Russia}



\author{X. Lu}
\affiliation{Beijing National Laboratory for Condensed Matter Physics, Institute of Physics, Chinese Academy of Sciences, Beijing 100190, China}

\author{H. Luo}
\affiliation{Beijing National Laboratory for Condensed Matter Physics, Institute of Physics, Chinese Academy of Sciences, Beijing 100190, China}


\author{A. N. Vasiliev}
\affiliation{M.V. Lomonosov Moscow State University, Faculty of Physics, 119991 Moscow, Russia}
\affiliation{National Research South Ural State University, Chelyabinsk 454080, Russia}
\affiliation{National University of Science and Technology (MISiS), Moscow 119049, Russia}

\author{V. M. Pudalov}
\affiliation{P.N. Lebedev Physical Institute, Russian Academy of Sciences, 119991 Moscow, Russia}

\author{Xiao-Jia Chen}
\affiliation{Center for High Pressure Science and Technology Advanced Research, Shanghai, 201203, China}

\author{Mahmoud Abdel-Hafiez}
\email{mahmoudhafiez@gmail.com}
\affiliation{National University of Science and Technology (MISiS), Moscow 119049, Russia}
\affiliation{Center for High Pressure Science and Technology Advanced Research, Beijing, 100094, China}

\date{\today}

\begin{abstract}
We report on the Andreev spectroscopy and specific heat of high-quality single crystals BaFe$_{1.9}$Ni$_{0.1}$As$_{2}$. The intrinsic multiple Andreev reflection spectroscopy reveals two anisotropic superconducting gaps $\Delta_L \approx 3.2 \textendash 4.5$\,meV, $\Delta_S \approx 1.2 \textendash 1.6$\,meV (the ranges correspond to the minimum and maximum value of the coupling energy in the $k_xk_y$-plane). The $25 \textendash 30 \%$
anisotropy shows the absence of nodes in the superconducting gaps. Using a two-band model with $s$-wave-like gaps $\Delta_L \approx 3.2$\,meV and $\Delta_S \approx 1.6$\,meV, the temperature dependence of the electronic specific heat can be well described. A linear magnetic field dependence of the low-temperature specific heat offers a further support of $s$-wave type of the order parameter. We find that a $d$-wave or single-gap BCS theory under the weak-coupling approach cannot describe our experiments.

\end{abstract}


\maketitle
\section{INTRODUCTION}

One of the crucial issues to elucidate the mechanism leading to high-temperature superconductivity is the nature of pairing, e.g., the symmetry and structure of the superconducting order parameter. Conventional phonon-mediated superconductors and unconventional cuprate superconductors \cite{Cup,SC} are commonly characterized by distinct $s$-wave and $d$-wave pairing symmetries with isotropic nodeless and anisotropic nodal gap distributions, respectively. In conventional superconductors, the electron-phonon interaction gives rise to the attraction between electrons near the Fermi-surface (FS) with opposite momenta and opposite spin directions, which eventually forms Cooper pairs. In the $d$-wave pairing symmetry, the order parameter changes sign in the basal plane, forcing the gap to vanish to zero along diagonal directions ($k_{y}$=$\pm k_{x}$) \cite{J}. Although there is a general consensus in the theory that electron-electron interactions play an important role in the formation of Cooper pairs in cuprates as well as in pnictides \cite{Hir,d,mait,Hir2}, many aspects such as the role of orbital fluctuations, magnetism, the mechanism of chemical tuning, the role of the vicinity of the Fermi level to the band edge to , and the resultant pairing symmetry remain unsettled \cite{SC,Hir,Nat1,Nat2}. Experimental \cite{Hardy,ARPES,STM,R,Tan1,Mah18} and theoretical \cite{Lee,RT} studies of pnictides, particularly the 122 system, show that the superconducting (SC) gap structure is not universal and differs in various materials.

In multiband materials there are obvious reasons for changing the character of interactions with doping, due to the electron-hole asymmetry, i.e. the difference in the effective masses and sizes of the hole-like bands at the $\Gamma$ point and of the electron-like bands at the $M$ point. The importance of the FS proximity to nesting conditions and the role of spin fluctuations in the pairing mechanism therefore change as relative sizes of the electron and hole bands vary with doping. For example, in the framework of the $s^{++}$ model, the competition between spin-fluctuation-mediated repulsion and attraction via orbital fluctuations may cause gap anisotropy or nodes \cite{Sai}. Even for the most intensively studied Ba$_{1-x}$K$_{x}$Fe$_{2}$As$_{2}$ system, where the majority of experimental data unambiguously shows the nodeless $s$-type gap, there still are several experimental indications for the pairing symmetry to be nontrivial. Indeed, in the inelastic neutron scattering experiments \cite{castellan_PRL_107.177003} it was noted that the magnetic excitations spectrum splits into two incommensurate peaks because the growing mismatch in the hole and electron FS-sheets accompanies with the fall of $T_c$ with hole doping. The latter is consistent with $s^{\pm}$-symmetry pairing calculations. In the phase sensitive Josephson tunneling measurements \cite{burmistrova_PRB_2015}, $s^{\pm}$ symmetry was found for current injected in the $ab$-plane and $s_{++}$ -- for current injected along $c$. The $s^{\pm}$-symmetry reveals itself also through the second harmonic oscillations of the Josephson current as a function of the RF-power \cite{burmistrova_PRB_2015}.

However, KFe$_{2}$As$_{2}$ shows signs of $d$-wave superconductivity, the issue that is still highly debated among the community. For instance, the BCS-ratios obtained for the KFe$_{2}$As$_{2}$ system \cite{Mah2} are comparable with the two-band $s$-wave fit of the penetration depth data for K-122 (1.28 and 5.31 for the small and large gaps, respectively) and do not exceed the corresponding values for the isomorphic compound RbFe$_{2}$As$_{2}$ (1.74 and 5.7) \cite{Mah3}. This similarity might reflect the presence of nodes in the SC order parameter of KFe$_{2}$As$_{2}$. This is also supported by the specific heat data reported for K$_{1-x}$Na$_{x}$Fe$_{2}$As$_{2}$, in which the obtained $T^{2}$ and $\sqrt{H}$ behavior of the specific heat gives an evidence for the line nodes \cite{Mah}. Additionally, from penetration depth \cite{hashimoto_PRB_82_014526 2010} and from heat conduction measurements with KFe$_{2}$As$_{2}$ \cite{J}, the line nodes were reported in the energy gap on the large zone-centered hole sheets. For BaFe$_{2}$(As$_{0.7}$P$_{0.3}$)$_{2}$ compound of the same Ba-122 family, in angle-resolved photoemission spectroscopy measurements \cite{zhang_NatPhys_2012} line nodes were found for the $\alpha$ condensate. For NaFe$_{1-x}$Co$_{x}$As compound, from London penetration depth measurements \cite{cho-tanatar_PRB.86.020508} the deviation of the superfluid density from the $T^2$ dependence has been prescribed to the presence of the line nodes. The line nodes were concluded to be significant at the dome edges though were also reported for the optimal doping. For Ba(Fe$_{1-x}$Co$_{x}$)$_{2}$As$_{2}$ (0.038 $< x <$ 0.127) in thermal conductivity measurements \cite{reid_PRB 82.064501(2010)} the nodes were reported for heat flow along the $c$ axis; the nodes are accidental and appear as $x$ deviates from the optimal doping. In IR measurements \cite{fischer_100.0692}, the $T^2$ temperature dependence of the penetration depth at the lowest temperatures has been interpreted as manifestation of the nodes in the gap function.

Here we present study of the superconducting order parameter in nearly optimal BaFe$_{2-x}$Ni$_{x}$As$_{2}$ system using the bulk probe (specific heat) and the direct local probe (intrinsic multiple Andreev reflection effect (IMARE) spectroscopy). Data measured by both techniques show the presence of two nodeless gaps, whereas the IMARE study resolved a moderate in-plane anisotropy of both gaps ($\sim 25 \textendash 30\%$).

\section{SAMPLES AND MEASUREMENT TECHNIQUES}

BaFe$_{2-x}$Ni$_{x}$As$_{2}$ single crystals were grown by the FeAs self-flux method. Details for the growth process and sample characterization were published elsewhere \cite{S2,PRB}. The Ni-doping levels reported throughout the paper refer to the actual Ni-content that was found to be 80\%
of the nominal level $x$ through the inductively coupled plasma analysis of the as-grown single crystals. The low temperature specific heat down to 0.4\,K and the resistivity up to 9\,T were measured with the Physical Property Measurement System (PPMS) using the adiabatic thermal relaxation technique.

In order to determine directly the value and the symmetry of the SC order parameter for $x = 0.1$ sample, we used multiple Andreev reflection effect (MARE) spectroscopy of superconductor - constriction - superconductor (ScS) junctions formed using the break-junction technique \cite{Moreland,BJ}. The crystal was prepared as thin plate with $a \times b \times c = (2 \textendash 4)\times (1 \textendash 2) \times (0.05 \textendash 0.2)$\,mm$^{3}$, and mounted onto a springy sample holder so that the crystal $ab$-plane was always perpendicular to force applied to the center of the holder. We attached the sample with four In-Ga pads in order to enable 4-probe measurements, and then cooled down the sample holder with the sample to $T$ = 4.2\,K. Next, the holder was precisely curved, which caused a crack in the single crystal splitting it in two parts. Under such deformation, two weakly connected SC banks were generally formed. In the used set-up, the microcrack is located deep in the bulk and is remote from current leads, therefore, the ScS region is protected from contaminations and overheating during the experiment \cite{BJ}.

Like in any layered material, the cleaved surface of BaFe$_{2-x}$Ni$_{x}$As$_{2}$ single crystal shows steps and terraces along the $ab$-planes. The steps often lead to natural ScSc-$\dots$-S stack structures, where the SC Fe$_{2-x}$Ni$_x$As blocks act as ``S'', and the metallic Ba layers play a role of constriction \cite{Ba}. Due to uniformity of ``S'' and ``c'' bulk areas in the crystal structure, the constrictions become highly transparent (more than 95\%),
and act as thin normal metal, constituting a natural chain of identical SnS junctions \cite{BJ,Ba}. Gently adjusting the holder curvature, it is possible to finely tune the constriction dimension and resistance, the number of contacts in the array is accidental. For the stacks in the above mentioned geometry, current always flows along the $c$-direction. As a part of the pristine crystal structure, the ScSc-$\dots$-S array consists of clean constrictions protected from degradation. Earlier we showed \cite{EPL,BJ} that the larger number of junctions in the stack $m$, the greater is the contribution of the $bulk$ to the dynamic conductance of such array. In particular, for the Ba-122 family, although the surface states seem significant \cite{vanHeumen}, their influence may be neglected when studying the array contact. As a result, the natural ScSc-$\dots$-S array surpass single junction or artificial mesa structure in terms of quality, cleanliness, sharpness of the resonant features in dI-dV-curves, and heat sink.


\begin{figure*}
\includegraphics[width=20pc]{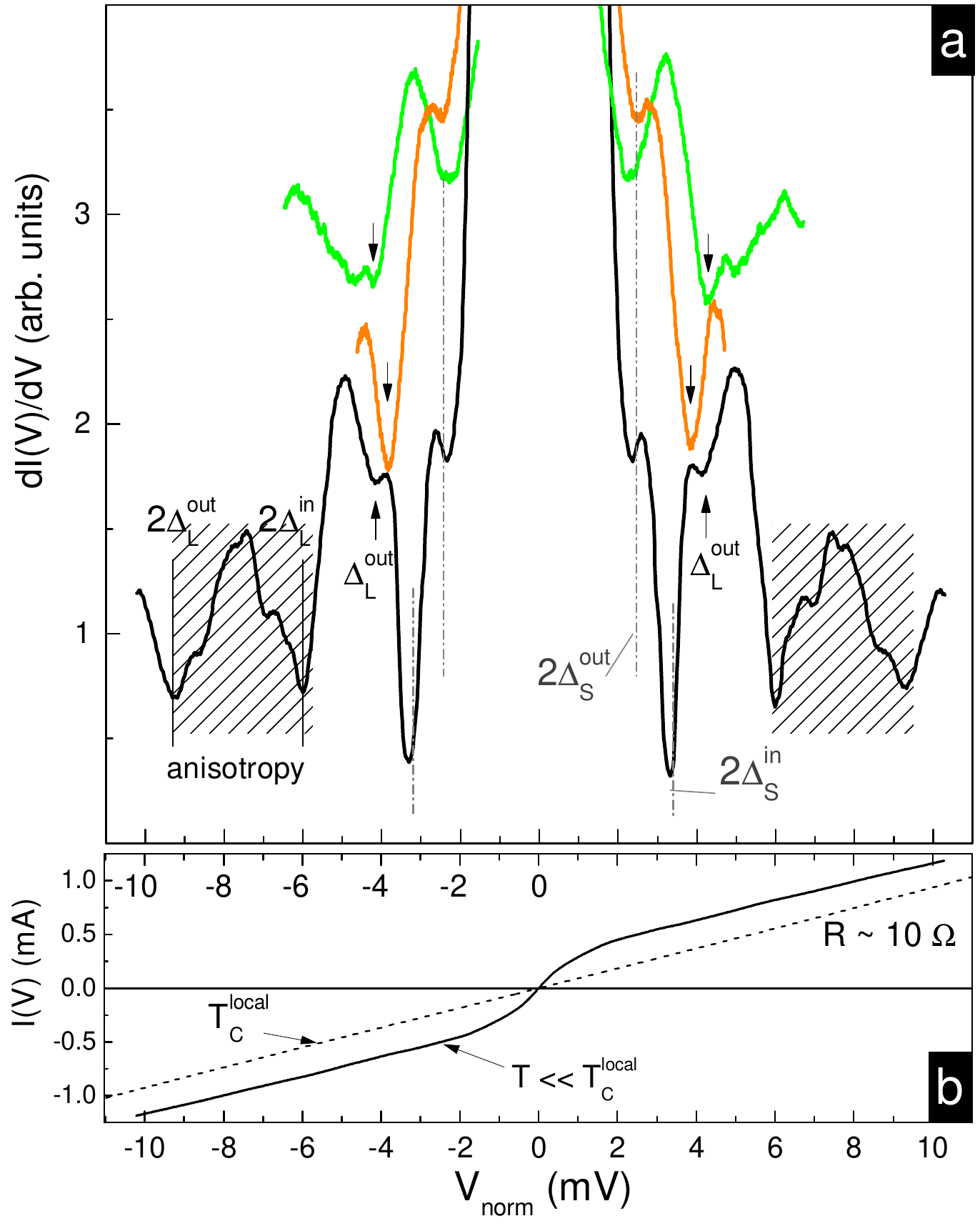}
\includegraphics[width=20pc]{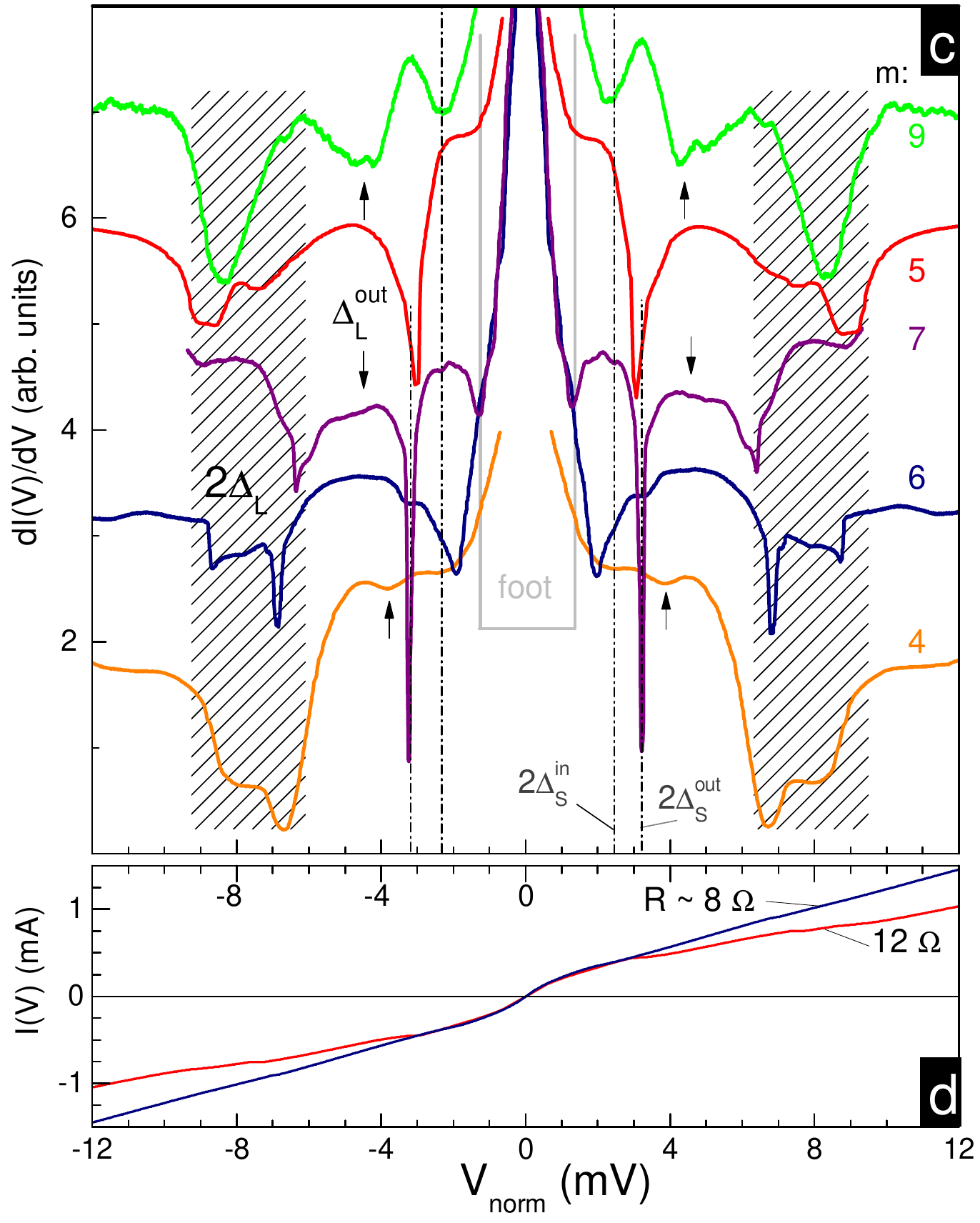}
\caption{\textbf{a}: Dynamic conductance spectrum (black line) of $m = 5$ junction Andreev array measured at $T = 4.2$\,K. For the large gap edges $\Delta_L^{out}$ and $\Delta_L^{in}$, the position of the doublet-like Andreev features $n_L = 1$ is marked by solid lines, the anisotropy range \textemdash by shaded areas; the second ($n_L = 2$) harmonic of the $\Delta_L^{out}$ \textemdash by black arrows. The doublets corresponding to the small gap edges $\Delta_S^{out}$ and $\Delta_S^{in}$ ($n_S = 1$), are shown by dash-dot vertical lines. Here ``out'' and ``in'' indexes relate to the outer and inner edges of the in-plane gap angular distribution in the $k$-space, respectively. \textbf{b}: Current-voltage characteristics measured at $T = 4.2$\,K (solid line), and at $T_C^{local} \approx 18.5$\,K (dashed line), for the contacts shown in \textbf{a}. \textbf{c}: Dynamic conductance spectra (the curves are offset vertically for clarity; the absolute ordinate has no physical meaning here). \textbf{d} Current-voltage characteristics of various Andreev arrays measured at $T = 4.2$\,K (colors correspond to those on the panel \textbf{c}). The notations are the same as in \textbf{a}. The low-bias fragments (zoomed vertically) of the $m = 4$ and 9 junction arrays are shown in the panel \textbf{a} in order to demonstrate the second subharmonic of the large gap. $\Delta_L(\theta) \approx 3.2\textendash4.5$\,meV ($\approx$ 30\%
in-plane anisotropy), $\Delta_S(\theta) \approx 1.2\textendash1.6$\,meV ($\approx$ 25\%
in-plane anisotropy). The $V$ axes for $I(V)$ and $dI(V)/dV$ curves are normalized to those of a single SnS junction.}
\end{figure*}

\section{RESULTS}
\subsection{Intrinsic Multiple Andreev Reflection Effect (IMARE) Spectroscopy}

\begin{figure*}[tbp]
\includegraphics[width=15pc]{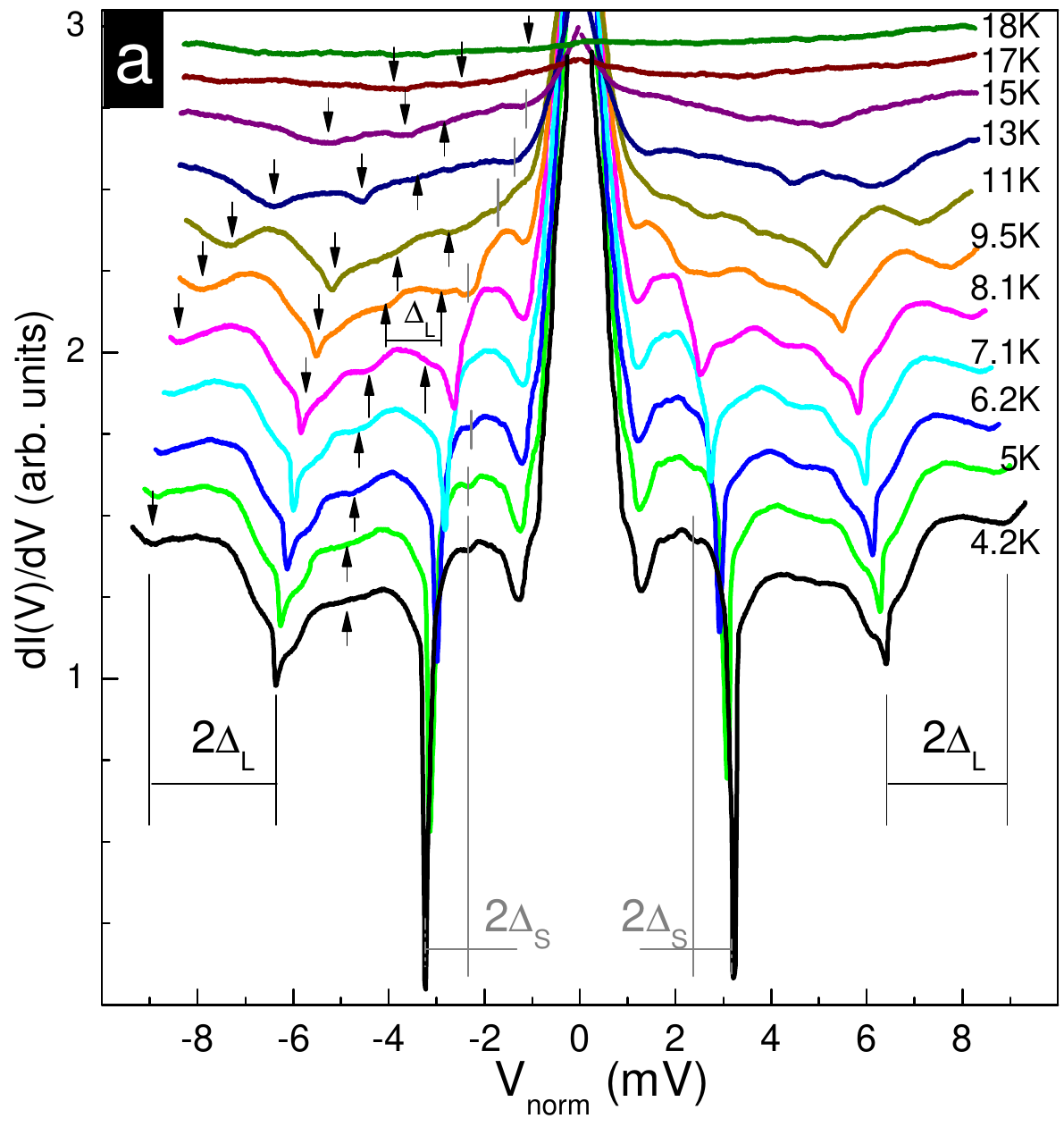}
\includegraphics[width=7pc]{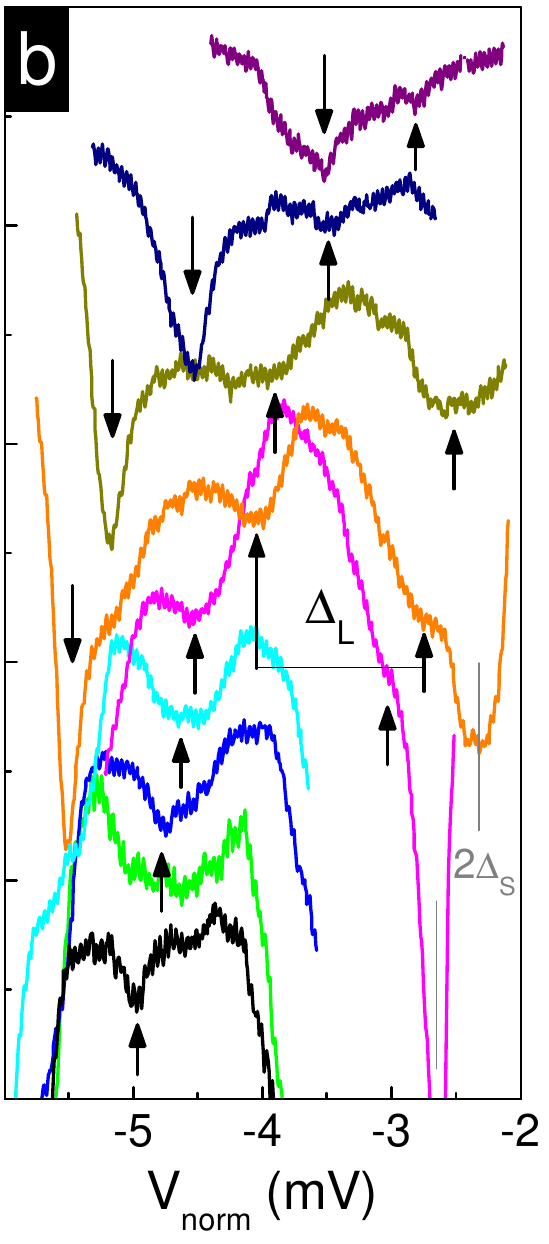}
\includegraphics[width=16pc]{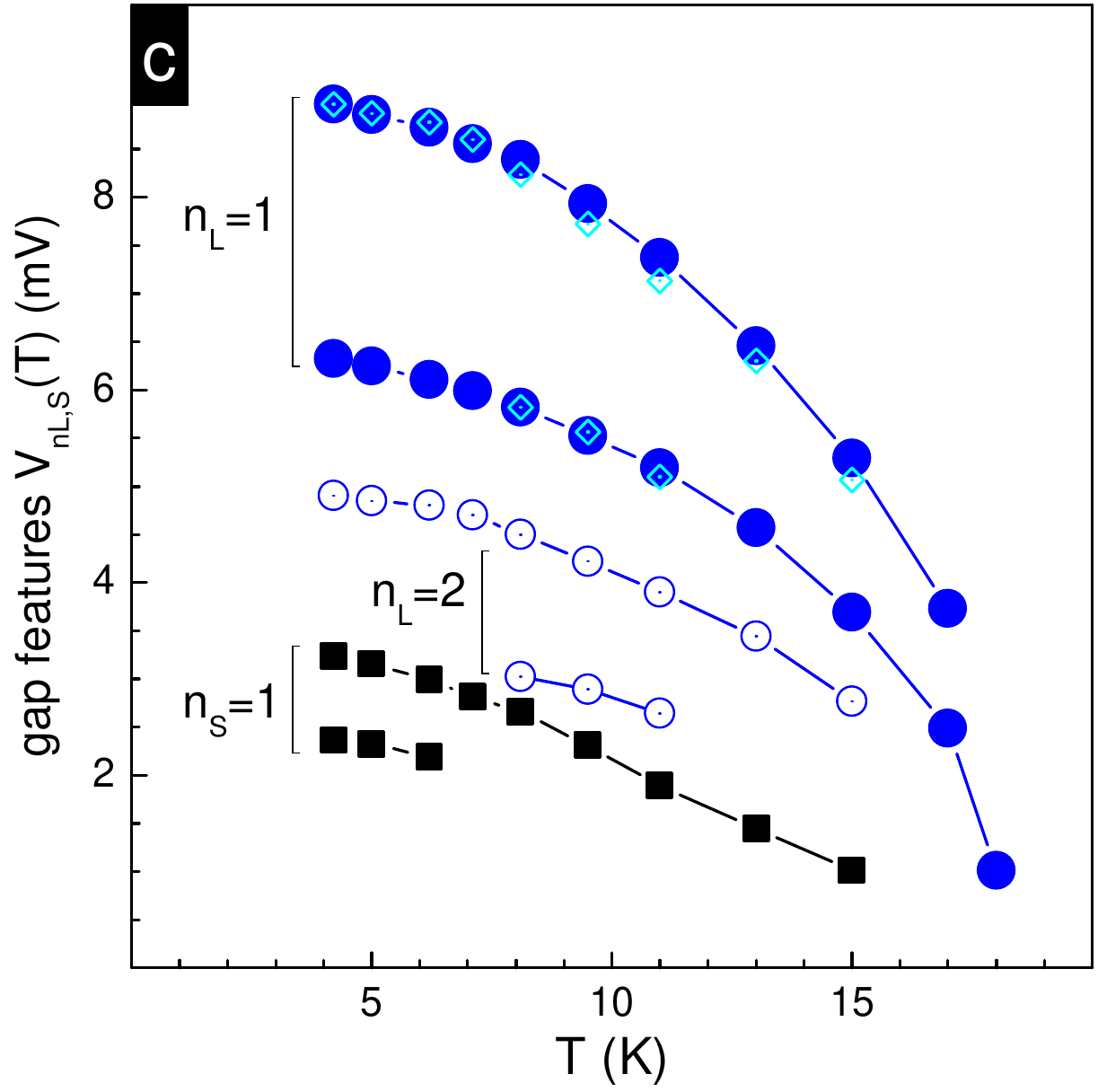}
\caption{\label{fig:wide} \textbf{a:} Normalized dynamic conductance spectrum of Andreev array measured at $T = 4.2 \textendash 18$\,K. The $dI(V)/dV$ curves are offset vertically for clarity. The positions of the first-order doublet-like Andreev dips for the large gap ($n_L=1$, $2\Delta_L$ labels) are marked by down arrows; the second-order dips ($n_L=2$, $\Delta_L$ label) are shown by up arrows. The features for the small gap are labeled as $2\Delta_S$ and gray solid lines. \textbf{b:} Vertically zoomed fragments of the spectra shown in \textbf{a} containing the $n_L = 2$ features, the linear background is suppressed for clarity. The notations are similar to \textbf{a}. \textbf{c:} Temperature dependence of the positions of the first ($n_L=1$, solid circles) and the second ($n_L = 2$, open circles) Andreev dips of the $\Delta_L^{ex,in}$, and the main dips ($n_S=1$, squares) of the $\Delta_S^{ex,in}$. The normalized dependence $V_{n_L=2}(T) \times 2$ is shown by rhombs for comparison with the $V_{n_L=1}$ (solid circles).}
\end{figure*}

The ScS break junctions demonstrated multiple Andreev reflection effect (MARE), similar to that in the high-transparent SnS contact (where $n$ is thin layer of normal metal) with Andreev transport \cite{Sharvin,Andreev}. MARE manifests in the $excess$ conductance at any voltages which raises significantly at low bias (so called foot area). Apart from the foot, a series of dynamic conductance features called subharmonic gap structure (SGS) appears at positions $V_{n} = 2\Delta/en$ ($n$ = 1, 2, $\dots$ is natural number) \cite{OTBK,Arnold,Averin,Kummel,Devereaux}. In particular, the first-order ($n = 1$) feature is located at $2\Delta/e$ bias. This simple formula directly associates the SC gap value with the location of the Andreev features at any temperatures up to $T_c$ \cite{OTBK,Kummel} and provides direct measurement of the gap temperature dependence $\Delta(T)$. Tracing the $\Delta$ vanishing to zero with temperature enables also to determine the local temperature $T_c^{local}$ of the contact area transition to normal state.

For the high-transparent (95 \textendash 98 \%)
metallic constriction (typical for our break junctions), the SGS shows a series of dynamic conductance dips for both nodeless and nodal gap \cite{Devereaux,Kjaergaard}. The coexistence of two SC gaps would cause, obviously, two SGS's in the $dI/dV$-spectrum of SnS junction. For the $dI(V)/dV$ spectrum of SnSn-$\dots$-S Andreev array containing $m$ SnS junctions ($m$ is natural number), the position of features caused by \textit{bulk} properties of material scales by a factor of $m$. Hereafter, the SGS caused by the intrinsic multiple Andreev reflections effect (IMARE), appears at positions

{\begin{equation}
V_n^i = \frac{2\Delta_i}{en} \times m,~~~~~ m, n = 1, 2, \dots
\end{equation}}
where $i =$ L(S) for the large (small) gap, respectively.

The IMARE resembles the intrinsic Josephson effect \cite{PonIJE,Moll}, and was firstly observed in Bi-family cuprates \cite{PonIMARE}, and then in other layered high-temperature superconductors \cite{SSC2012,EPL,Ba}. Although the formed array could contain accidental number of junctions, the actual $m$ value is uniquely determined by finding such minimum natural number, which provides a good scaling of all features for array to collapse onto those for a single contact. The IMARE spectroscopy provides direct local probe (measured within the contact area with diameter about dozens nm) of \textit{bulk} parameters of the superconductor. This advantage provides accurate determination of the gap, whereas the high quality of the break junctions facilitates high spectra resolution and, hence, a possibility to study the fine structure of the $dI(V)/dV$ spectrum \cite{EPL,BJ}.

The $k$-space angular distribution of the gap value strongly affects the shape of the SGS dips (see the Appendix for details). In case of an isotropic gap, the SGS minima are well-pronounced and symmetric, whereas nodal gap (such as ``fully anisotropic'' $s$-wave or $d$-wave) manifests strongly suppressed and asymmetric minima \cite{BJ,Devereaux,Cuevas}. As for extended $s$-wave symmetry without nodes, each subharmonic reveals itself in the $dI(V)/dV$ as a doublet of two coupled dips, whose positions determine the outer and inner edges of the gap angular distribution in the momentum space \cite{Li2013,BJ,Ba}. In the used set-up, the carriers with various momenta in the $k_xk_y$-plane pass through the constriction along the $c$-direction, with the velocity component $v_c \ll v_a, v_b$.
Therefore, this technique enables to observe the $k_xk_y$-plane anisotropy solely \cite{BJ}.

The CVC and dynamic conductance spectra typical for Andreev array in BaFe$_{1.9}$Ni$_{0.1}$As$_2$ are shown in Figs. 1a,b. At $T = 4.2$\,K, the $I(V)$ curve demonstrates a notable foot, with the conductivity at low biases $\approx 4.3$ times larger than that in the vicinity of the SC transition (at $T_c^{local} \approx 18.5$\,K), exceeding the conventional $11/3 \approx 3.7$ value \cite{OTBK} and evidencing for the multiple Andreev reflections in high-transparency constriction. Additionally, in order to to check the actual regime (ballistic or diffusive), we take the normal-state bulk resistivity $\rho(22{\rm K}) \approx 1.9 \times 10^{-4}\, \Omega \cdot {\rm cm}$ for the sample under study \cite{Chen}, the typical normal resistance per one contact $R \sim 10$\,$\Omega$ (see Fig.~1b), and use Sharvin formula $a = \sqrt{\frac{4}{3\pi} \frac{\rho l}{R}}$ (where $2a$ is the contact diameter, $l$ is the elastic mean free path) \cite{Sharvin}. Accounting the product of bulk resistivity and the carrier mean free path $\rho l = 1.65 \times 10^{-13}$\,$\Omega \cdot$ m$^2$ for Co-doped Ba-122 \cite{Machida}, we roughly estimate the elastic carrier mean free path $l^{el} \approx 87$\,nm. In turn, the constriction dimension $a \approx 84$\,nm is of the same order of magnitude as $l^{el}$. Strictly speaking, in case the contact diameter is much less than $\sqrt{l^{el}l^{in}}$, where $l^{in}$ is inelastic scattering length (diffusive regime), MAR still occur, although the spectroscopic signal is reduced \cite{PCSbook}. In this case, the high-order subharmonics are suppressed, nonetheless, the position of the first gap feature remains $2\Delta/e$ \cite{Kummel}. Typically, the energy relaxation length $l^{in} > 100 l^{el}$, limiting the point contact diameter to $2a < 10 l^{el} \sim 0.9$\,${\rm \mu m}$ in the diffusive regime. Therefore, in the break-junctions formed in BaFe$_{1.9}$Ni$_{0.1}$As$_2$, the diffusive regime is likely realized, with $1-2$ subharmonics expected in the dI(V)/dV spectrum. The $V$ axis of the dynamic conductance spectrum shown in Fig.~1a was scaled down to a single SnS junction by a factor of $m = 5$. At bias voltages $\pm 6.4$ and $\pm 9$\,mV, the $dI(V)/dV$ spectrum demonstrates the first-order ($n_L = 1$) doublet minima of the large gap edges $\Delta_L^{in}$ and $\Delta_L^{out}$, respectively (emphasized in Fig.~1a,c by shaded areas). Using Eq. (1), we determine the outer and inner gap edges $\Delta_L^{out} \approx 4.5$\,meV, $\Delta_L^{in} \approx 3.2$\,meV, respectively. The doublet shape of these features points to a moderate gap anisotropy in the $k$-space $1 - \Delta_L^{in}/\Delta_L^{out} \approx 30\%$.
The second order subharmonic $n_L = 2$ of the outer gap edge is also clear and marked by the black vertical arrows. As for $n_L = 2$ for the inner edge of the large gap expected at $V \approx \pm 3.2$\,mV, it is overlapped with the small gap SGS. The notable minima at $\pm 3.2$\,mV are more intensive than those for the large gap, and do not satisfy Eq. (1) as the third subharmonic of $\Delta_L$, therefore we attribute them to the small superconducting gap $\Delta_S$. With such interpretation, the doublet Andreev features observed as sharp intensive dips at $\pm 3.2$\,mV, and minor dips at $\pm 2.4$\,mV, determine $\Delta_S^{out} \approx 1.6$\,meV and $\Delta_S^{in} \approx 1.2$\,meV, respectively. The latter values signify $\sim 25\%$
in-plane anisotropy of the small gap. The higher-order subharmonics of the small gap are poorly visible. Possible reasons of this might be: (i) overlapping of the subharmonics with a quickly rising foot, and/or (ii) a shorter mean free path in the current channel for the bands with $\Delta_S$. The determined moderate anisotropy $\approx 25 \textendash 30 \%$
of both gaps points at the absence of nodes in the  $\Delta_{L,S}(\theta)$ angular distribution. Noteworthily, the obtained values do not depend on the contact resistance and on the number of junctions in the Andreev array, therefore, they should be attributed to two bulk order parameters.

A set of dynamic conductance spectra of Andreev arrays at $T = 4.2$\,K is shown in Fig.~1b. The arrays were formed in one and the same crystal under a mechanical readjustment, the corresponding $m$ are shown next to each curve. The dI(V)/dV curves were offset vertically for clarity and normalized to a single junction. The absolute normal resistance of arrays (not shown in Fig.~1c) depends on the diameter of column (usually $10-100$\,nm) and is proportional to the contact number $m$. We have not succeeded in producing the single contacts with the studied Ba-122 samples. The $m$ numbers for each array were estimated from the comparison of spectra for all the arrays obtained and taking into account the BCS-ratio in the weak coupling limit 3.5 ($2\Delta_L / k_B T_c > 3.53$).

After such scaling, the position of all the gap features remains constant for all the spectra, regardless of the contact dimension and resistance, thus providing an evidence for their \textit{bulk} nature. Note, that the fact that the spectrum may be scaled by integer numbers $m$ signifies the approximate equivalence of the junctions in the array. The doublet features of the large and the small gap are characteristic for the dynamic conductance spectra of Andreev arrays measured for BaFe$_{1.9}$Ni$_{0.1}$As$_2$. The lower panel of Fig.~1b shows normalized I(V) curves for $m = 5$ and 6 junction arrays, with pronounced foot at low bias voltages. The beginning of the foot is also marked in the upper panel of Fig.~1b as the exponential rise of the dynamic conductance at the corresponding dI(V)/dV. The second subharmonic of the large gap at position $eV = \Delta_L^{out}$ is also resolved in some spectra; in order to demonstrate it, we show the low-bias fragments (zoomed vertically) of $m = 4$ and 9 junction arrays in Fig.~1a.

\begin{figure}[tbp]
\includegraphics[width=20pc,clip]{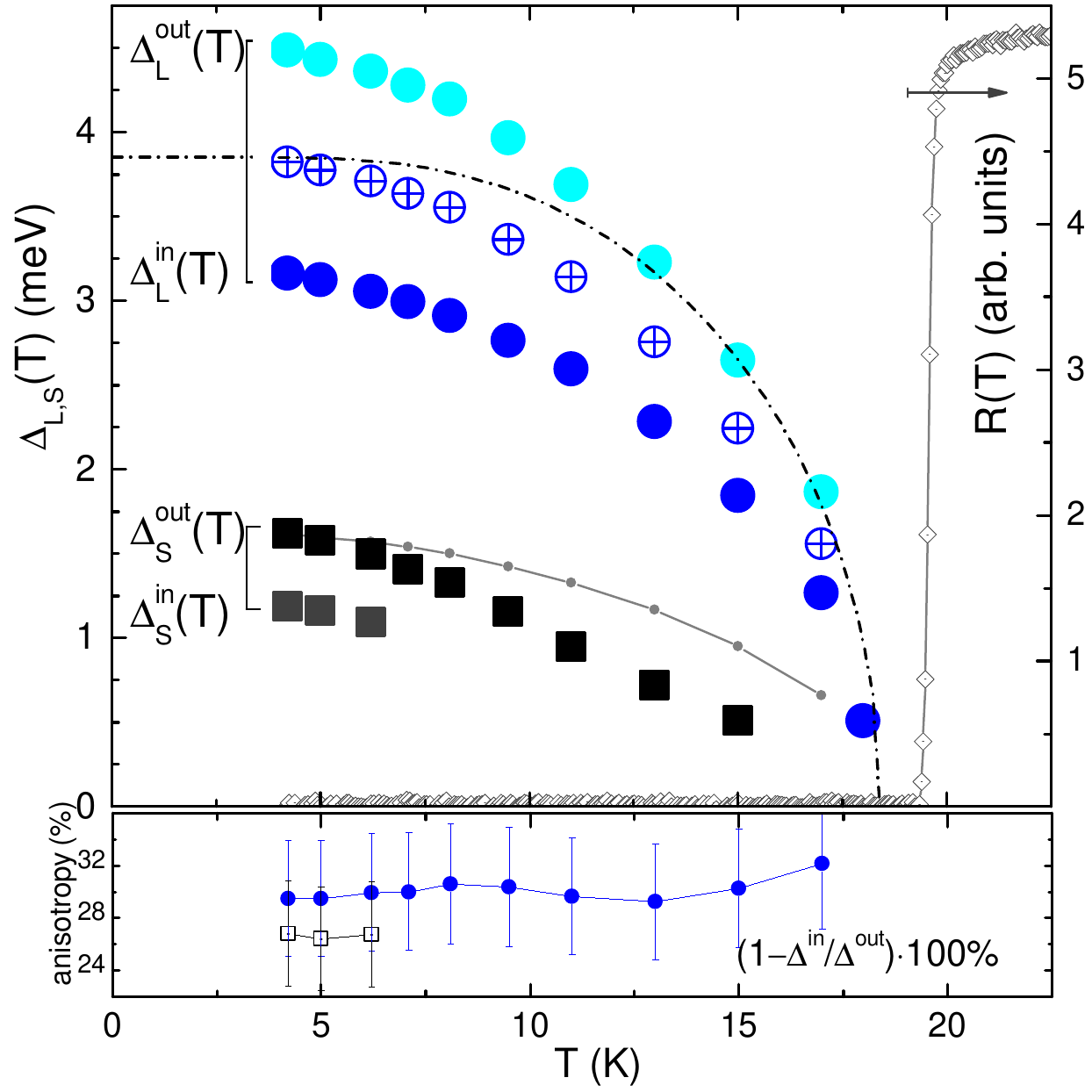}
\caption{\textbf{Upper panel:} temperature dependence of the superconducting gaps in BaFe$_{1.9}$Ni$_{0.1}$As$_2$ (using the data of Fig.\,2c). For the large gap, the extreme values $\Delta_L^{out}(T)$ and $\Delta_L^{in}(T)$ are shown by solid circles, the mean value $\Delta_L = (\Delta_L^{out}+\Delta_L^{in})/2$ \textemdash by crossed circles. The small gap $\Delta_S^{out}(T)$ is shown by solid squares, the normalized dependence $\Delta_L(T) \cdot \Delta_S^{out}(0)/\Delta_L(0)$ is shown for comparison by gray line with dots. Dash-dot line is a single-band BCS-like curve, open rhombs show the bulk resistive transition. \textbf{Lower panel:} temperature dependence of the anisotropy of the large gap (circles) and the small gap (squares) determined as $(1 - \Delta_i^{in}(T)/\Delta_i^{out}(T)) \cdot 100\%$}
\end{figure}

Temperature evolution of the dynamic conductance is shown in Fig.~2\textbf{a}. The bottom $dI(V)/dV$ spectrum in Fig.~1\textbf{b} was measured within the temperature range from 4.2\,K to 18\,K. With temperature increase, the $2\Delta_L$ and $2\Delta_S$ features gradually turn to zero. At 18\,K, the spectrum becomes almost flat, which signifies the vicinity of the $T_c^{local}$. Simultaneously, the conductance of the contact decreases with temperature increasing in accordance with the predictions \cite{Klapwijk} for Andreev regime. In Fig.~2\textbf{a}, for clarity, the dynamic conductance curves were offset vertically. The evolution of the gap features with temperature is shown in Fig.\,2\textbf{a} by down arrows ($2\Delta_L/e(T)$), up arrows ($\Delta_L/e(T)$), and gray solid bars ($2\Delta_S/e(T)$), and in Fig.~2\textbf{c} by solid circles, squares, and open circles, respectively. The first-order features of $\Delta_L^{out}$, $\Delta_L^{in}$, and $\Delta_S^{out}$ are clearly seen till $T_c^{local}$. The minor $\Delta_S^{in}$ feature is resolved at low temperatures, until smeared at $\approx 6$\,K. As for the second $\Delta_L$ subharmonic (detailed in Fig.~2\textbf{b}), its outer feature is observed at low temperatures solely, while the inner one merges with the sharp $2\Delta_S^{out}$ dip. As the small gap decreases more rapidly with temperature, at $T \approx 8$\,K its inner feature gets resolved, so that the whole $n_L = 2$ doublet becomes visible. Being multiplied by a factor of two, the temperature dependence of the position of the $n_L = 2$ subharmonic (rhombs in Fig. 2\textbf{c}), matches the first one ($n_L = 1$, solid circles), thus confirming these features to belong to one and the same SGS of the large gap.

Temperature dependences of the large gap edges (solid circles), the small gap edges (open circles), and the mean magnitude of the large gap $\Delta_L = (\Delta_L^{out} + \Delta_L^{in})/2$ (crossed circles), are shown in Fig.\,3. Dash-dot line corresponds to a single-band BCS-like curve, which, obviously, do not fit the experimental $\Delta_L(T)$. The $\Delta_L(T)$ curve slightly bends down as compared with the BCS-like function. The small gap decreases almost linearly in the interval for $5-15$\,K. In Fig. 4, we also present the normalized $\Delta_L(T) \cdot \Delta_S^{out}(0)/\Delta_L(0)$ data (gray line with dots) showing the difference between the large and the small gap temperature dependences. The outer and inner $\Delta_L$ dips behave similarly, which allows to attribute them to one and the same but anisotropic SC order parameter. The different temperature behavior of the $\Delta_S$ features evidences their relation to another distinct SC order parameter. Both gaps turn to zero at the common critical temperature $T_C^{local} \approx 18.5$\,K, almost coinciding with the beginning of the superconducting $R(T)$ transition of the bulk crystal (rhombs).

The lower panel of Fig.\,3 shows temperature dependence of the gap anisotropy. The large gap anisotropy remains within $29 \textendash 32\%$
independent on temperature. To obtain similar dependence for the small gap, further studies are required due to the vanishing intensity of $\Delta_S^{in}$ Andreev features. Nonetheless, at low temperatures, the small gap anisotropy was estimated as $26 \textendash 27\%$.

\begin{figure}[tbp]
\includegraphics[width=20pc,clip]{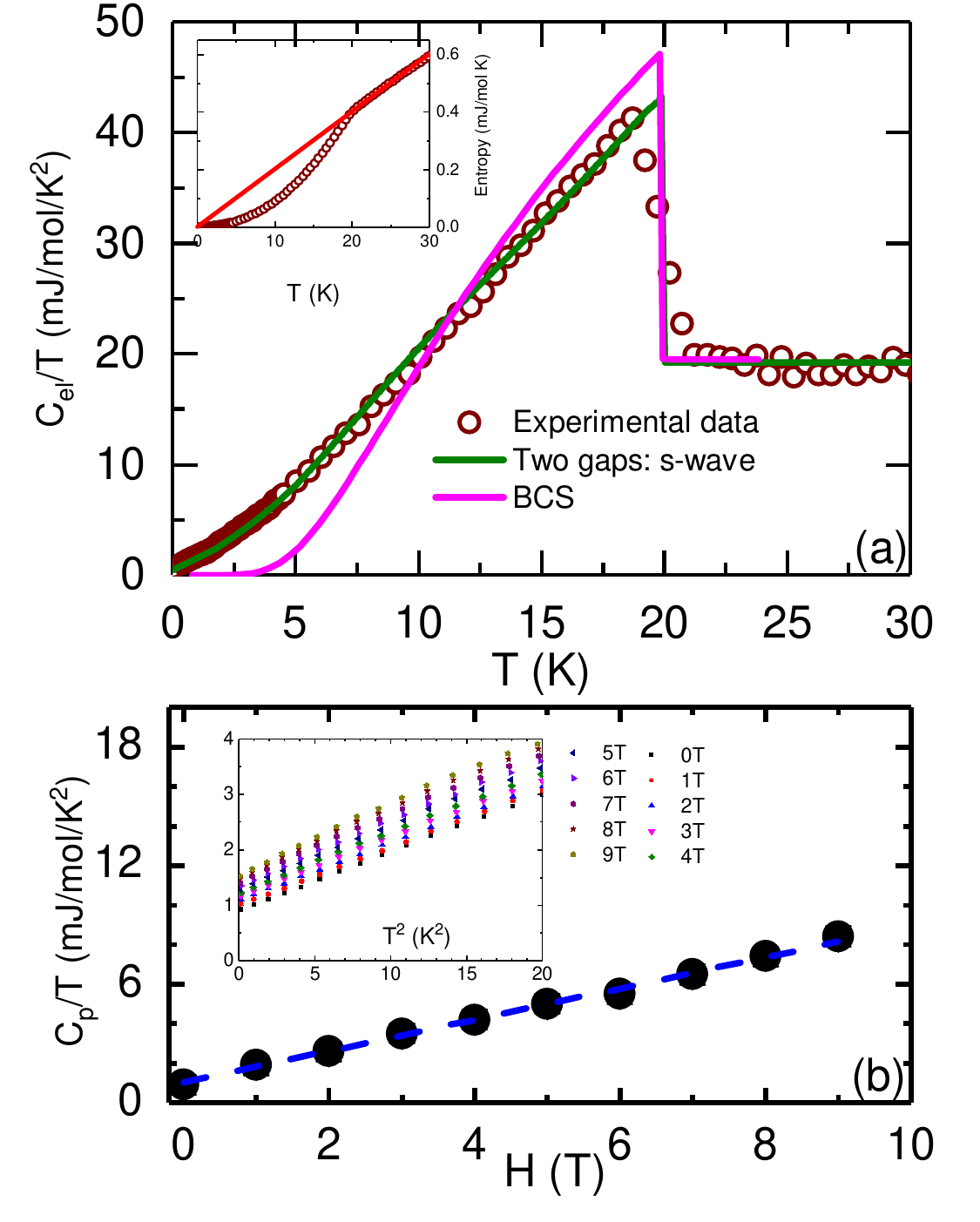}
\caption{\textbf{a:} temperature dependence of the electronic specific heat in BaFe$_{1.9}$Ni$_{0.1}$As$_2$. The inset presents the entropy in the normal and superconducting state as a function of temperature. \textbf{b:} The  field  dependence  of  the  mixed  state  quasiparticle  contribution for $H \parallel c$. The dashed lines represent the phenomenological linear fit. The inset shows the specific heat plotted as $C_{p}/T$ vs. $T^{2}$ measured under various magnetic fields up to 9\,T in the low temperature region.}
\end{figure}

\subsection{Specific heat}
Specific heat also provides a probe for the symmetry and structure of the SC order parameter. In order to determine the specific heat related to the SC phase transition we need to separate the phonon ($C_\mathrm{ph}$) and electron ($C_\mathrm{el}$) contributions. First we address the zero-field $T$-dependence of the electronic specific heat data plotted as $C_{el}/T$ vs $T$ (main panel of Fig.\,4\textbf{a}). A clear sharp jump is observed, which is due to the SC phase transition. In order to determine the specific heat related to the SC phase transition we need to estimate the $C_\mathrm{ph}$ and $C_\mathrm{el}$ contributions to $C_p$ in the normal state. In order to determine the phononic contribution to the specific heat for $x$ = 0.25, the following relation is used: $C^{x=0.25}_{ph} = C^{x=0.25}_{tot} - C^{x=0.25}_{el} $, where $C_{el}^{x=0.25}$ is {just} $\gamma T$. The same shape of the phononic heat capacity in the SC samples ($x = 0.1$) and overdoped sample ($x = 0.25$) is assumed. Therefore, the specific heat of the SC samples can be represented by:
\begin{equation}\label{eq1}
C_{el}^{SC}/T = C_{tot}^{SC}/T - \emph{g} C_{ph}^{x=0.25}/T,
\end{equation}
which allows us to calculate the $C_\textup{el}$ of the SC samples. The small deviation of the scaling factor \emph{g} from unity,  plausibly related to experimental uncertainties, demonstrates that the above procedure represents a very good method to determine the phonon background. The value of $g$ was determined from the requirement of equality between the normal and SC state entropies at $T_\textup{c}$, {that is} $\int_{0}^{T_\textup{c}}\left(C_\textup{el}/T\right)dT = \gamma_\textup{n} T_\textup{c}$, where $\gamma_\textup{n}$ is the normal state electronic specific heat coefficient. We started with $g= 1$, but we found that the entropy conservation criterion is satisfied with {$g = 0.95$}. Physically, this indicates that the substitution of Fe by Ni does not substantially affect the lattice properties~\cite{PRB}. The almost linear temperature dependence of $C_\mathrm{el}/T$ of the SC samples indicates that the specific heat data cannot be described by a single BCS gap. In order to illustrate this we show a theoretical BCS curve with $\Delta = 1.764\, k_{\mathrm{B}}T_\mathrm{c} = 2.23$~meV in Fig.\,4\textbf{a}. One can see that systematic deviations from the data are observed in the whole temperature range below $T_\mathrm{c}$. This clearly indicates that the gap structure of our systems is more likely to be nodeless $s$-wave. The jump height of the specific heat at $T_c$ is found to be $\Delta C_{el}/T_{c}$ $\approx$ 24(1)~mJ/mol K$^2$. From our determined $\gamma_n$ values, we estimate the universal parameter $\Delta C_{el}/\gamma_nT_c$ = 1.25. the value, however, is lower than of the BCS weak coupling approximation of 1.43. This points toward a multiband (gap) scenario with $s$-, $p$-, or $d$-wave pairing. Since a single-gap scenario cannot describe our data, we applied a phenomenological two-gap model in line with multigap superconductivity reported by various experimental and theoretical studies on different compounds within the FeAs family~\cite{m1,m2,m3,m4,m6}. We analyzed our data utilizing the generalized $\alpha$ model, which has been proposed to account for the thermodynamic properties of multiband, multigap superconductors like, e.g., MgB$_{2}$~\cite{Bouquet2001}:

\begin{equation}\label{eq2}
    \frac{S}{\gamma_{n}T_{c}}=-\frac{6\Delta_{0}}{\pi^{2}k_{B}T_{c}}\int_{0}^{\infty}[f\ln f+ (1-f)\ln (1-f)]dy,
\end{equation}
\begin{equation}\label{eq3}
     \frac{C}{\gamma_{n}T_{c}}= t\frac{d(
     \frac{S}{\gamma_{n}T_{S}})}{dt},
\end{equation}
where $f$ = $[$exp($\beta E$) + 1]$^{-1}$, $\beta$ = ($k_{B}T_c$)$^{-1}$ and the energy of the quasiparticles is given by $E$ = $[\epsilon^{2} + \Delta^{2}(t)]^{0.5}$ with $\epsilon$ being the energy of the normal electrons relative to the Fermi surface. The integration variable is y = $\epsilon$/$\Delta_0$. In Eq.~\ref{eq3} the scaled gap $\alpha = \Delta_0/k_BT$ is the only adjustable fitting parameter. The temperature dependence of the gap is determined by $\Delta(t) = \Delta_0\delta(t)$, where $\delta(t)$ is obtained from the table in Ref.~\cite{muehlschlegel59}. In case of two gaps the
thermodynamic properties are obtained as the sum of the contributions from the two gaps, i.e., $\alpha_1$ = $\Delta_1(0)/k_BT_c$ and $\alpha_2$ = $\Delta_2(0)/k_BT_c$ with their respective weights $\gamma_1/\gamma_n$ and $\gamma_2/\gamma_n$.

To calculate the theoretical curves C$_{el}/\gamma_nT$ the parameters $\Delta_1$, $\Delta_2$, their respective ratios $\gamma_1$ and $\gamma_2$ and the ratio $\gamma_{r}/\gamma_{n}$ are left free for fitting ($\gamma_{r}$ represents the small residual value of the non-superconducting electrons of our sample at low temperatures). We note that $C_\mathrm{el}/T$ almost saturates at low temperature; however, it does not extrapolate to zero, yielding a residual electronic specific-heat value $\gamma_\textup{r}$ = 1.6\,mJ/mol K$^{2}$ for the investigated system. The finite value of $\gamma_\textup{r}$ indicates a finite electronic density of states at low energy, even in zero applied field. We mention that the presence of a finite $\gamma_\textup{r}$ is common in both electron- and hole-doped 122 crystals and that the value of $\gamma_\textup{r}$ in our case is remarkably low, showing the good quality of our investigated single crystals. However, the origin of this residual term is still unclear. It may be because of an incomplete transition to the SC state or because of broken pairs caused by disorder or impurities in unconventional superconductors, and/or spin-glass behavior. The best description of the experimental data is obtained using values of $\Delta$$_{S}$ = 1.6~meV and $\Delta$$_{L}$ = 3.2~meV. The calculated specific heat data are represented by the solid line in Fig.4\textbf{a}.

Next we discuss the field dependence of specific heat through the vortex excitation in the mixed state, which is another independent test sensitive to the gap structure. It has been well demonstrated that for the isotropic $s$-wave, $\gamma(H) \propto H$ because the specific heat in the vortex state is dominated by the contribution from the localized quasiparticle in the vortex core~\cite{Vol}. Recently, Storey {\it et al.}~\cite{JGS} pointed out that the number of Caroli-de Genned bound states increases linearly with the field due to the linear increase in the number of vortices entering the sample. On the other hand, for the line nodes $\gamma(H) \propto H^{0.5}$, the quasiparticle contributing to the density of states come from regions away from vortex core, close to the nodes and the supercurrents around a vortex core in the mixed state cause a Doppler shift of the quasiparticle excitation spectrum~\cite{GEV}. We plotted the field dependence of specific heat coefficient in the main panel of Fig.\,4\textbf{b}. Obviously, for the investigated system, $C_\mathrm{p}/T$ varies almost linear with magnetic field (Fig.\,4\textbf{b}). The magnetic field enhances the low-temperature specific heat continuously, indicating the increase of the quasiparticle density of states at the Fermi surface (see the inset of Fig.\,4\textbf{b}). The roughly linear magnetic field dependence of the specific heat suggests that at least one (dominating) SC condensate is fully gapped, most likely related to the hole pocket. It should be noted that the slight curve bending indicates the presence of at least two gaps. In order to further verify this point, the low $T$ specific heat data in the mixed state have been analyzed in more details, see above.

\section{Discussion}

Deriving a solid picture of the SC gap symmetry in the Fe-based superconductors constitutes a challenge. This is due to the wide diversity of experimental results for the gap structure, due to indirect character of the experimental probes, and lack of the true bulk probes. For instance, surface probes, such as ARPES \cite{1,2,3,4,5,6,Evtushinsky,Aswartham} show two full gaps with no nodes, however, the situation is less clear for bulk probe \cite{7,8,9,10} measurements which draw a more diverse picture pointing to a non-universal gap-structure. However, the vast majority of bulk probes gives consistent results for the same family members of the Fe-based superconductors and it was pointed out that one can explain the discrepancies between the data obtained with bulk and surface probes by surface type mechanisms such as surface electronic reconstruction or surface depairing \cite{Hir}.

Using two different probes, specific-heat and break-junction, with the same samples, we investigated the SC order parameters in BaFe$_{1.9}$Ni$_{0.1}$As$_{2}$ system. The orthorhombic distortion and the superconductivity are intimately coupled in BaFe$_{2-x}$Ni$_{x}$As$_{2}$ \cite{N}. Theoretically \cite{N2}, it is well established that when both the SC and the orthorhombic order parameters are taken into account, the anisotropy of the SC coherence length is enhanced. Similar to the recent work on Ba$_{1-x}$K$_{x}$Fe$_{2}$As$_{2}$ \cite{N3}, this approach could also explain the appearance of the gap anisotropy in  BaFe$_{2-x}$Ni$_{x}$As$_{2}$.

Firstly, the temperature dependence of $C_\mathrm{el}$/$T$ data in BaFe$_{1.9}$Ni$_{0.1}$As$_{2}$ cannot be described by a single BCS gap. Rather, the $C_{el}/T$ data may be well fitted with the two-band model assuming the two nodeless gaps. Secondly, $C_\mathrm{p}/T$ varies roughly linearly with magnetic field, which suggests that at least one (dominating) SC condensate is fully gapped, which is probably related to the hole pocket. The two above facts suggest that the  gap structure  for our system  most likely consists of two SC condensates with at least one of them (dominating) being of the nodeless $s$-wave type symmetry. In addition, the slight $C_p(T)$ curve bending at low-temperatures indicates the presence of at least two gaps.

Strictly speaking, the IMARE spectroscopy provides a direct probe of the extremal gap value at $T < T_c$, and, in general, unable to specify its belonging to a certain band, or phase. However, summarizing the IMARE data shown in Figs. 1-3, one can draw several indirect conclusions. We resolved the four extremal gap values, $\Delta_S^{in} \approx 1.2$\,meV, $\Delta_S^{out} \approx 1.6$\,meV, $\Delta_L^{in} \approx 3.2$\,meV, and $\Delta_L^{out} \approx 4.5$\,meV (at $T = 4.2$\,K), which demonstrate two typical temperature dependences (see Fig. 3). Since the majority of ARPES probes reported two distinct SC gaps in Ba-122 family compounds \cite{Evtushinsky,Aswartham}, we attribute the 1.2 and 1.6\,meV values to the SC condensate with the small gap, and 3.2 and 4.5\,meV values \textemdash for the condensate with $\Delta_L$. The maximum BCS-ratios for the large gap $2\Delta_L^{out}/k_BT_c^{local} \approx 5.6$ and for the small gap $\Delta_S^{out}/k_BT_c^{local} \approx 2$ are in excellent agreement with the ARPES data \cite{Evtushinsky} for Ca$_{1-x}$Na$_x$Fe$_2$As$_2$.
The BCS-ratio exceeds the weak-coupling BCS-limit 3.5, thus pointing to a strong effective coupling within the bands where the $\Delta_L$ order parameter is developed. For the small gap, $2\Delta_S/k_BT_c^{local} \approx 1.5 \textendash 2 \ll 3.5$ is caused by a nonzero interband coupling.

The shape of the Andreev doublet dips observed in the dI(V)/dV spectra shown in Fig. 1, is in qualitative agreement with an extended $s$-wave symmetry without nodes (see the Appendix). With temperature, the two dips of the doublet for the large gap behave in a same way. The above mentioned results evidences for a moderate anisotropy of both gaps $\approx 25 \textendash 30\%$
in the basal plane. Nonetheless, in the spectra shown in Fig. 1, the outer dip of the small gap dominates over the inner one. This may indicate a nonuniform gap spectral weight in the basal plane, with a prevalence of $\Delta_S \approx 1.6$\,meV. The specific heat measurement also supports this conclusion, taking into account the values $\Delta_L = 3.2$\,meV, $\Delta_S = 1.6$\,meV which dominate in the bulk properties and give the best fit of the $C_{el}(T)/T$ data.

The determined BCS-ratios agree well with those determined in our earlier IMARE studies of Ba$_{0.65}$K$_{0.35}$Fe$_{2}$As$_{2}$ \cite{Ba}, as well as with some ARPES \cite{Evtushinsky,Aswartham}, $H_{c1}$ \cite{Ba,Ren}, and specific heat data \cite{Hard,Pramanik,Hardy2} for Ba-122 family compounds of various composition. Remarkably, the anisotropy range determined in the present study of BaFe$_{1.9}$Ni$_{0.1}$As$_2$ almost coincides with that reported by IMARE for the (Ba,K) compound \cite{Ba}.


\section{Conclusions}

We reported experimental data on the gap structure and the anisotropic superconducting properties of BaFe$_{1.9}$Ni$_{0.1}$As$_{2}$. Specific heat in zero field follows a two-band model with $s$-wave type order parameter. In magnetic field, $C/T$ develops  linearly with magnetic field. Additionally, intrinsic multiple Andreev reflection effect (IMARE) spectroscopy data resolved substantial anisotropy of both superconducting gaps. The SC gap values in the $k_xk_y$ plane are $\Delta_L \approx 3.2 \textendash 4.5$\,meV, $\Delta_S \approx 1.2 \textendash 1.6$\,meV; this data may be considered as evidence of an extended $s$-wave symmetry with $\approx 25-30 \%$
in-plane anisotropy. Both used techniques show the absence of nodes in the superconducting gaps. The BCS-ratio estimated for the large gap $2\Delta_L/k_BT_c^{local} = 4.0 \textendash 5.6$ exceeds the weak-coupling BCS-limit and shows a strong intraband coupling.

We acknowledge G. Volovik,  A. Bianconi, and G. Karapetrov for discussions. The work was supported by RSF (16-12-10507). S.A.K. acknowledges the Russian Foundation for Basic Research (project no. 17-02-00805a). The measurements were partly performed using equipments of the Shared facility center at LPI. This work has been supported by the Ministry of Education and Science of the Russian Federation in the framework of Increase Competitiveness Program of NUST (MISiS) grant K2-2017-084, by acts 211 of the Government of Russian Federation, Contracts No. 02.A03.21.0004 and 02.A03.21.0011.

\section{Appendix}
\subsection{Influence of the gap anisotropy to Andreev spectrum}

Consider SnS junction formed in two-band superconductor with two distinct order parameters $\Delta_L$ and $\Delta_S$ coexisting in the momentum space. In fully ballistic SnS junction, since elastic process, MAR does not cause interband scattering. The Andreev conductivity therefore involves two parallel channels \cite{BJ,PRB2017}. As a result, in the superconductor with $\Delta_L > 3\Delta_S$, the dynamic conductance spectrum would show two separate SGS, one corresponding to the $\Delta_L$ at large bias voltages, and another to the $\Delta_S$ at low bias, at any temperatures up to the $T_c$. In opposite (inelastic) case, a strong interband scattering would mix the channels in $k$-space, with an appearance of additional SGS at positions $eV = (\Delta_L + \Delta_S)/n$. Nonetheless, our experimental data with two-gap superconductors with almost isotropic order parameter, such as Mg$_{1-x}$Al$_x$B$_2$ and the 1111 family oxypnictides \cite{SSC2004,BJ,SSC2012,EPL,PRB2017} do not show such combined SGS. Noteworthily, such SGS was not observed even in pure MgB$_2$ crystals with structural defects and low mean free path \cite{SSC2004,BJ}. In case of momentum-dependent SC order parameter, the dynamic conductance of SnS Andreev contact (as well as of other types of tunneling junction) can show complex and nontrivial features. However, analyzing the shape of Andreev features, it is possible to distinguish between several basic symmetries. The two extremal gap values resolved in the dI(V)/dV spectra (see Figs. 1,~2) can originate from either $k$-space anisotropy of the SC order parameter or two independent SC gaps. In order to simulate the shape of Andreev features for the cases considered in Fig. 6, we used the raw spectra calculated in \cite{Devereaux} for the  $s$ and $d$ symmetries. Note the spectra in Fig.~5 simulate the shape of Andreev features barely, the exponential background leading to minima distortion was suppressed. The Andreev transport component in case of isotropic $s$-wave gap (Fig. 5, curve no. 1) results in sharp and well-pronounced dI(V)/dV feature. Such symmetric minimum of a finite width \cite{Devereaux} was used in the hereafter simulations. For all the resulting curves, we took the two extremal gap values, so that the dI(V)/dV dips appear at the same positions $eV/2\Delta = $ 1 and 0.7 ($30\%$ splitting).

\begin{figure}
\includegraphics[width=20pc,clip]{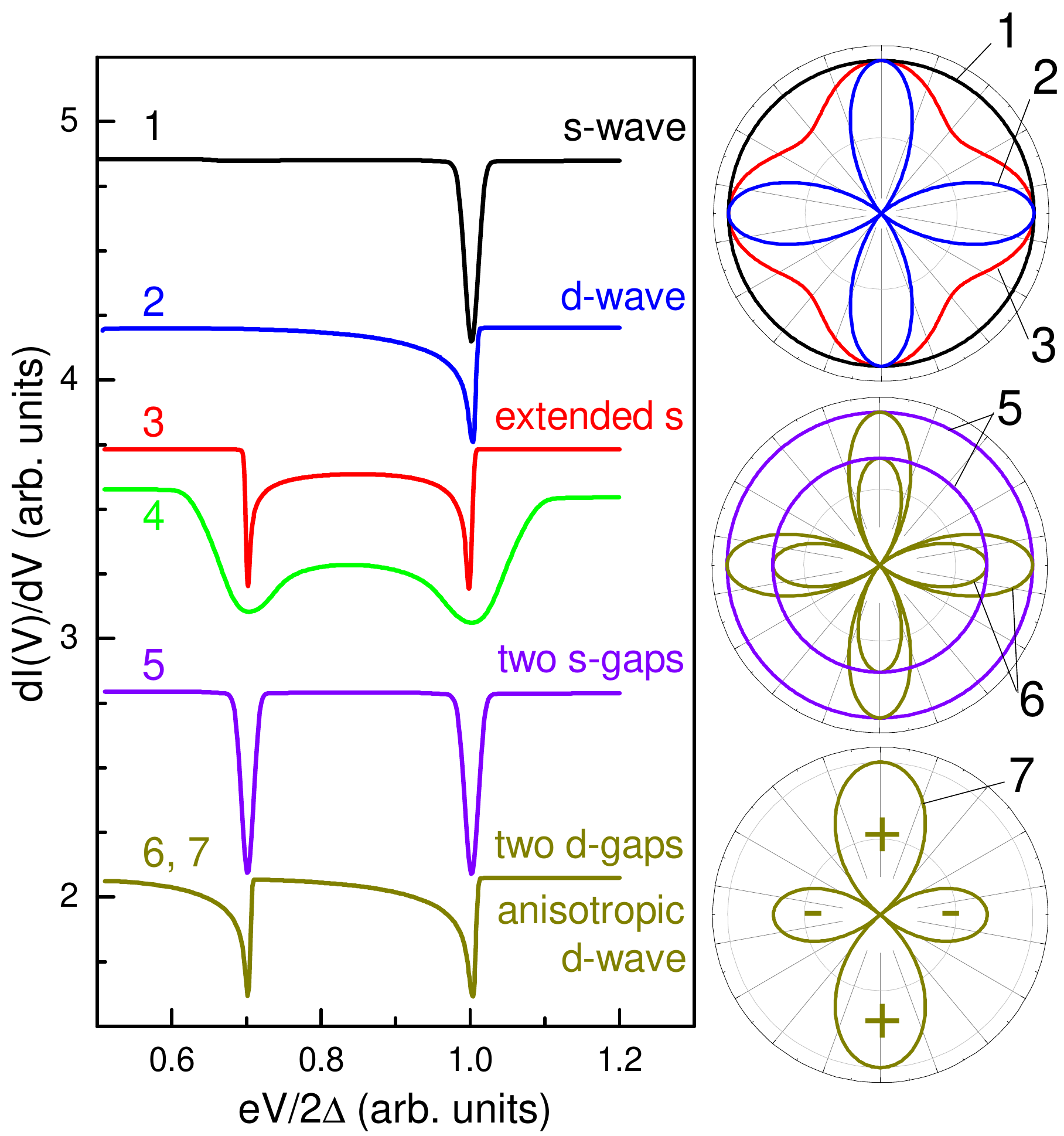}
\caption{Typical shape of the SnS-Andreev feature (left panel) for the various symmetries of the gap in the $k_xk_y$-plane shown in the right. For a single SC order parameter: (1) $s$-wave, (2) $d$-wave, (3) extended $s$-wave ($\Delta(\theta) \propto cos(4\theta)$) with $\Delta^{out} = 1$, $\Delta^{in} = 0.7$ and hence $30\%$
anisotropy in the $k$-space; the dI(V)/dV no. (4) is similar to (3) but for the gap with a distorted $cos(4\theta)$ angle distribution. For two distinct order parameters: (5) two $s$-wave gaps $\Delta_1 = 1$, $\Delta_2 = 0.7$; (6) two $d$-wave gaps. A distorted $d$-wave gap with the angular distribution $\Delta(\theta) = 0.85cos(2\theta) + 0.15$, the Andreev feature looks as (6) as well. The curves are offset vertically for clarity, the exponential background typical for MAR is suppressed.}
\end{figure}

A $d$-wave gap ((2), the blue line) causes asymmetric and less intensive dips (for the corresponding dI(V)/dV, the amplitude was gained by a factor of 10). Similar shape of the dip is typical for a sign-preserved nodal gap. To simulate a nodeless extended $s$-wave symmetry (red line), a simple $\Delta(\theta) \propto cos(4\theta)$ angle distribution was taken, with $\Delta^{out} = 1$, $\Delta^{in} = 0.7$ (curve no. 3 in Fig. 5). Such anisotropic gap causes Andreev doublet of two mirrored asymmetric dips and an arch between them. More complex $\Delta(\theta)$ dependence would entangle the shape of the doublet, for example, a distorted $\cos(4\theta)$ symmetry, smears the minima and lowers the connecting arch (curve no. 4). Anyhow, characteristic of anisotropic but nodeless gap distribution in the momentum space is the couple which never holds out the background of the $dI(V)/dV$ spectrum.

The two coexisting SC order parameters $\Delta_1$ and $\Delta_2$ would cause two overlapping subharmonic gap structures (SGS) in the dI(V)/dV (curves nos. 5, 6). Each of these SGS's consists of a single dip, however, due to the chosen close gap values $\Delta_1 = 1$ and $\Delta_2 = 0.7$, the resulting Andreev features resemble doublet as well. For isotropic $s$-wave gaps (violet lines), the dips are symmetric (curve no. 5), whereas two $d$-wave gaps (yellow lines) cause asymmetric minima (curve no. 6). For the latter case, with bias voltage decrease, the dynamic conductance drops down abruptly (specifying the gap value), then saturates until reaches the background of the spectrum. Noteworthily, the distorted $d$-wave gap with the angular distribution $\Delta(\theta) = 0.85cos(2\theta) + 0.15$ (with the amplitude of the positive and negative folds $\Delta^+ = 1$, $\Delta^- = 0.7$, correspondingly) causes the Andreev subharmonic like (6) as well. Nonetheless, the curve no. 6 strongly differs from the case of anisotropic but sign-preserved gap (no. 3,~4).

In the experimental spectra shown in Figs. 1,~2, the couple do not saturates enough to reach the background. Besides, doublets for the large gap show a fine structure along the couple, whereas the threshold dips look rather broaden as compared to those for the small gap. The latter, nonetheless, should not be attributed to poor experimental resolution, taking into account the sharp and well-pronounced Andreev dips for the small gap. Evidently, the observed shape of the $\Delta_L$ doublets resembles those in the curves nos. 3, 4, although its fine structure indicates the $\Delta(\theta)$ dependence differ from the simple $\cos(4\theta)$.

\end{document}